\documentclass[epj]{svjour}
\usepackage{graphics}
\usepackage{epsfig}

\newcommand{\dto}{D$_{2}$O}

\newcommand{\nue}{$\nu_{e}$}
\newcommand{\numu}{$\nu_{\mu}$}
\newcommand{\nutau}{$\nu_{\tau}$}
\newcommand{\nux}{$\nu_{x}$}

\newcommand{\teff}{$T_{\rm eff}$}
\newcommand{\costs}{$\cos\theta_{\odot}$}

\newcommand{\snoncfluxunc}{6.42^{+1.57}_{-1.57}\mbox{(stat.)}^{+0.55}_{-0.58}~\mbox{(syst.)}} 
\newcommand{\snomutauflux}{3.41^{+0.45}_{-0.45}\mbox{(stat.)}^{+0.48}_{-0.45}~\mbox{(syst.)}} 
\newcommand{\snoeflux}{1.76^{+0.05}_{-0.05}\mbox{(stat.)}^{+0.09}_{-0.09}~\mbox{(syst.)}}

\newcommand{\nsigmassno}{5.3}

\begin{document}
\title{Recent Results from the Sudbury Neutrino Observatory}
\author{A.W.P. Poon\inst{} for the SNO Collaboration
\thanks{This work is supported by Canada: NSERC, Industry Canada, NRC, Northern Ontario Heritage Fund, Inco, AECL, Ontario Power Generation, HPCVL, CFI; US: DoE; UK: PPARC.  We thank the SNO technical staff for their strong contributions}
}
%
%
\institute{Institute for Nuclear and Particle Astrophysics, Lawrence Berkeley National Laboratory, Berkeley, CA, USA}
\date{Received: date / Revised version: date}
%
\abstract{
The Sudbury Neutrino Observatory (SNO) measures both the flux of the electron-type neutrinos and the total flux of all active flavours of neutrinos originating from the Sun.  A model-independent test of neutrino flavour transformation was performed by comparing these two measurements.  In 2002, this flavour transformation was definitively demonstrated. In this talk, results from these measurements and the current status of the SNO detector  are  presented.
\PACS{
      {26.65.+t}{Solar neutrinos}    \and
      {14.60.Pq}{Neutrino mass and mixing} \and
      {95.85.Ry}{Neutrino, muon, pion, and other elementary particles; cosmic rays}
     } 
} 
\maketitle
\section{Introduction}
\label{intro}
For over three decades, solar neutrino
experiments ~\cite{bib:homestake}-\cite{bib:superk}
have been observing fewer neutrinos than what are predicted by the
detailed models (e.g. ~\cite{bib:bpb}) of the Sun.  These experiments probe different parts of the solar neutrino energy spectrum, and show an energy dependence in the measured solar neutrino flux.  These observations can be
explained if the solar models are incomplete or neutrinos undergo flavour transformation while in transit to the Earth. 
The Sudbury Neutrino Observatory, using 1000 tonnes of
99.92\% isotopically pure \dto\ as the neutrino target, was constructed to resolve this puzzle.  Detailed physical description of the SNO detector can be found in Ref.~\cite{bib:sno}.

The SNO detector detects solar neutrinos through the following channels:
\[\begin{array}{lcll}
    \nu_{e}+d & \rightarrow & p+p+e^{-} & \hspace{0.5in} \mbox{(CC)}\\ 
    \nu_{x}+d & \rightarrow & p+n+\nu_{x} & \hspace{0.5in} \mbox{(NC)} \\
    \nu_{x}+e^{-} & \rightarrow &  \nu_{x}+e^{-} & \hspace{0.5in} \mbox{(ES)} \\
\end{array}\]
The charged-current (CC) reaction on the deuteron is sensitive exclusively to \nue, 
and the neutral-current (NC) reaction has equal sensitivity to all 
active neutrino flavours (\nux ; $x=e,\mu,\tau$).  Elastic scattering 
(ES) on electron is also sensitive to all active flavours, but with 
reduced sensitivity to \numu\ and \nutau.  SNO is currently the only experiment that can simultaneously observe the {\it disappearance} of \nue's and their {\it appearance} as another active species.  By counting the free neutron in the final state of the NC reaction, the total active $^8$B neutrino flux can be inferred for neutrinos with energy above the 2.2-MeV kinematic threshold.   

In the first phase of the SNO experiment, the free neutron from the NC interaction was detected by observing the 6.25-MeV $\gamma$ ray following its capture on a deuteron.  To enhance the neutron detection efficiency, 2 tonnes of salt was added in the second phase of the experiment.  In this case, the free neutron is captured by $^{35}$Cl and a $\gamma$-ray cascade with a total energy of 8.6~MeV is emitted.  In the final phase of the experiment, $^3$He proportional counters, the ``Neutral-Current Detectors" (NCD)~\cite{bib:browne}, are deployed on a 1-m grid in the D$_2$O volume.  Free neutrons are captured by $^3$He in this phase.  Results~\cite{bib:snocc,bib:snonc,bib:snodn} from the first phase of SNO experiment and its prospects in the salt and NCD phases of the experimental program are presented in this talk.  

\section{Results from the Pure \dto\ Phase}
\label{sec:1}
The data presented in this talk were recorded between November 2,
1999 and May 28, 2001.   The target was pure \dto\ and over 450~million triggers were recorded.  The corresponding livetime was 306.4~days.

Detector diagnostic triggers and instrumental background events were first removed in the offline data processing.  The latter reduction was accomplished by a set of algorithms that  remove events that do not have the characteristics of Cherenkov light emission.  For the events that passed the first offline reduction, the calibrated times and positions of the hit photomultiplier tubes (PMTs) were used to reconstruct the vertex position and the direction of the particle.  The energy estimator then assigned an effective kinetic energy \teff\ to each event based on these reconstructed parameters and the number of hit PMTs.  The energy estimator used calibration results such as light attenuation as inputs, and was normalized to the detector response to a $^{16}$N source~\cite{bib:nsix}.  The linearity and spatial dependence of the energy estimator were checked by a $^{252}$Cf fission source and a 19.8-MeV $\gamma$-ray source~\cite{bib:poon}.  The neutron detection efficiency was calibrated with a $^{252}$Cf fission source.  

Radioactive decays of the daughters in the natural $^{232}$Th and $^{238}$U chains were the dominant backgrounds in the neutrino signal window.  The levels of  $^{232}$Th and $^{238}$U daughters in the detector were determined by radiochemical assays of the detector target~\cite{bib:ander} and by analyzing the Cherenkov signals from their decays in the neutrino data set.  The radioactive backgrounds were found to be better than those specified in the design criteria of the detector.  

\begin{figure}
\centering
\includegraphics[width=3.4in]{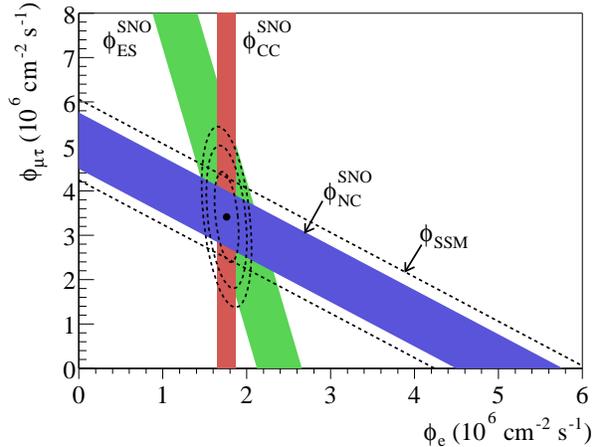}
\caption{Flux of solar \numu\ or \nutau\ {\it vs.} flux of \nue\ deduced from the three neutrino reactions in SNO.  The diagonal bands show the total $^{8}$B flux as predicted by the SSM~\cite{bib:bpb} (dashed lines) and that measured with the NC reaction (solid band).  The intercepts of these bands with the axes represent the $\pm 1\sigma$ errors.  The bands intersect at the fit values for $\phi_{e}$ and $\phi_{\mu\tau}$, indicating that the combined flux results are consistent with flavour transformation.}
\label{fig:1}     
\end{figure}

To test the null hypothesis of no neutrino flavour transformation in solar neutrinos, the extended maximum likelihood method was used to extract the CC,
ES and neutron (i.e. NC+background) contributions in the candidate data set.  Background contributions were constrained to the measured values.  Data distributions in \teff,  the volume-weighted radial 
variable $(R/R_{AV})^{3}$  and \costs\ were simultaneously fitted to the probability density functions (PDFs) 
generated from simulations.  $R_{AV}=600$~cm is the radius of the acrylic vessel, and $\theta_\odot$ is the angle between the reconstructed direction of the event and the instantaneous direction from the Sun to the Earth.   Assuming an undistorted $^{8}$B energy spectrum~\cite{bib:ortiz},  the fluxes of the \nue\ ($\phi_e$) and non-\nue\ ($\phi_{\mu\tau}$) components were found to be (in units of 10$^6$~cm$^{-2}$~s$^{-1}$):
\begin{eqnarray*}
\phi_{e} & = & \snoeflux \\
\phi_{\mu\tau} & = & \snomutauflux 
\end{eqnarray*}
Figure~\ref{fig:1} shows $\phi_{\mu\tau}$
{\it vs.} $\phi_e$ deduced from the SNO data.   Combining the statistical and systematic uncertainties in quadrature, $\phi_{\mu\tau}$ is \nsigmassno$\sigma$  above zero, and provides strong evidence for flavour transformation.  Removing the energy constraint, the signal decomposition was repeated using only the $\cos \theta_{\odot}$ and $(R/R_{\rm AV})^3$ distributions.  The total flux of active ${}^{8}$B neutrinos in the unconstrained fit was found to be (in units of 10$^6$~cm$^{-2}$~s$^{-1}$)
\begin{displaymath}
\phi_{\mbox{\tiny NC}}=  \snoncfluxunc 
\end{displaymath}
The correlation coefficients from the energy constrained and the unconstrained fits are shown in Table~\ref{tab:1}.

\begin{table}
\centering
\caption{Correlation coefficients in solar neutrino signal extraction in the pure D$_2$O phase}
\label{tab:1}       
\begin{tabular}{lcc}
\hline\noalign{\smallskip}
 & E constrained & E unconstrained  \\
\noalign{\smallskip}\hline\noalign{\smallskip}
NC, CC & -0.520 & -0.950 \\
CC, ES & -0.162 & -0.208 \\
ES, NC & -0.105 & -0.297 \\
\noalign{\smallskip}\hline
\end{tabular}
\end{table}

\section{Current Status of the Salt Phase}
\label{sec:2}
Neutrino flavour transformation can be an energy dependent process (e.g. the MSW mechanism~\cite{bib:msw}).  The energy constrained analysis in pure D$_2$O made the explicit assumption of an undistorted $^8$B neutrino spectrum.  The unconstrained results suffered from the strong anti-correlation between the CC and NC channels.  The addition of 2 tonnes of NaCl to the D$_2$O target was aimed at improving the separation between CC and NC without invoking the energy constraint.  Unlike the radiative capture of a neutron on deuteron, neutron capture on $^{35}$Cl typically produces multiple $\gamma$ rays while the CC and ES reactions produce single electrons. The greater isotropy of the Cherenkov light from neutron capture events relative to CC and ES events allows good statistical separation of the event types.  Simulations show that the statistical uncertainties in the NC channel reduce significantly in the unconstrained fit when Cherenkov light isotropy is included.  This is shown in Table~\ref{tab:2}.

There are a number of challenges to the analysis of the data in the salt phase.  A few of these challenges are described in the following.  To minimize the possibility of introducing biases, a blind analysis procedure is being used. The data set used during the development of the analysis procedures and the definition of parameters excluded an unknown fraction ($<$30\%) of the final data set, included an unknown admixture of muon-following neutron events, and included an unknown NC cross-section scaling factor.  At the time of this talk, the blindness constraints have not been removed.  

\begin{figure}
\centering
\includegraphics[height=2in]{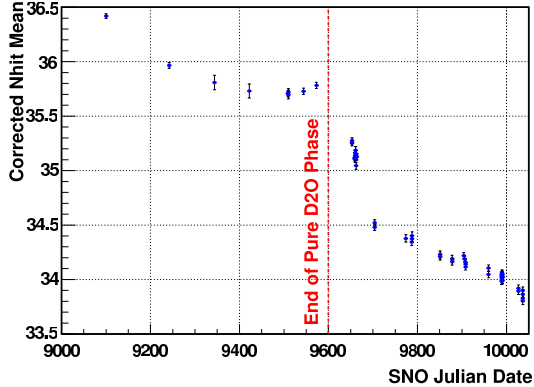}
\caption{Time evolution of the detector energy response to $^{16}$N calibration runs (6.13-MeV $\gamma$) taken at the center of the detector.  The number of fired PMTs, corrected for the variation of the number of online PMTs, is plotted against time.  The end of the D$_2$O phase is on May 28, 2001.}
\label{fig:2}     
\end{figure}

After the addition of NaCl into the D$_2$O target, the amount of detected Cherenkov light has dropped significantly over a short period of time.  Following this sharp drop, there has been a gradual decline in the detector gain (Figure~\ref{fig:2}).   There are a number of contributors to the drop in detector gain.  These include the degradation of the light concentrator surrounding each PMT, and the increase in the concentration of polymer colloids and Manganese due to the necessary modifications to the water purification procedures in the presence of NaCl.  Preliminary studies show that the absolutely energy scale uncertainty in the salt phase to be $\sim$1.1\%.  This is to be compared to 1.1\% in the pure D$_2$O phase.

The SNO simulation package, which uses EGS4~\cite{bib:egs} to simulate electromagnetic showers, predicted a more isotropic light distribution than what is observed from the energy and background calibration sources.  It was subsequently discovered that the electron scattering cross section in EGS4 neglects the Mott scattering terms, which reduce the amount of large angle scattering.  A correction derived from a Monte Carlo treatment of multiple scattering of electrons has been added to the SNO simulation package.  The agreement between the Monte Carlo prediction and the observed Cherenkov light isotropy improved significantly after this correction is made.  

$^{24}$Na is a non-negligible low energy background in the neutrino signal window.  This radioisotope is produced in the water circulation system by neutrons emitted from the rock wall in the underground laboratory.  It can also be produced {\em in-situ} by calibration sources (e.g. the $^{252}$Cf source).  In order to understand the production and the distribution of this background, a controlled neutron activation has been carried out.  The amplitude of this background in the neutrino signal window is found to be small.  

At the time of this talk, the SNO collaboration is finalizing the analysis procedures that are necessary for removing the data blindness constraints.    The results of the energy unconstrained NC measurement, along with its interpretation, will be reported in the near future.

\begin{table}
\centering
\caption{Expected statistical uncertainties in the salt phase}
\label{tab:2}       
\begin{tabular}{lccc}
\hline\noalign{\smallskip}
Fit Variables & CC & NC & ES \\
\noalign{\smallskip}\hline\noalign{\smallskip}
\teff, $(R/R_{AV})^{3}$, \costs & 4.2\% & 6.3\% & 10\% \\
\teff, $(R/R_{AV})^{3}$, \costs, Isotropy & 3.3\% & 4.6\% & 10\% \\
$(R/R_{AV})^{3}$, \costs, Isotropy & 3.8\% & 5.3\% & 10\% \\
\noalign{\smallskip}\hline
\end{tabular}
\end{table}

\section{Future Prospects and Summary}
\label{sec:3}

The SNO collaboration is now preparing for the NCD phase of its physics program.  In this phase, the NC neutrons are detected by the NCD array and the Cherenkov light from the CC signals are detected by the PMTs.  This separation of the detection mechanism will significantly reduce the statistical anti-correlation of the signals.  In addition, a measurement of the neutrino energy spectrum via the CC reaction can be made with higher precision with the reduction of NC signal in the Cherenkov light spectrum.  

Once the data blindness constraint is removed in the salt phase, the D$_2$O target will be desalinated.  All the $^3$He counters have been constructed and are being characterized in the underground laboratory.   There are intense activities in the final calibration of the detector in the salt phase, preparing for the desalination process, preparing for the NCD deployment, and integration of the NCD electronics and data acquisition system.  

The SNO experiment has demonstrated conclusively that solar $\nu_e$'s transform their flavour while in transit to the Earth.  A high precision measurement of the total active solar neutrino flux without any assumption of the energy dependence of the neutrino flavour transformation probability is near completion.  In the next few months, the SNO physics program will enter its NCD phase.  It is anticipated that the data from this NCD phase, along with data from the previous phases, will significantly enhance our understanding of the properties of neutrinos.

%
%

\end{document}